\documentstyle[fleqn]{article}

\newcommand {\tsub}[1]{_{\mbox{\protect\scriptsize #1}}}

\newcommand {\Ref}[1]{(\ref{#1})}
\newcommand
{\Stretch}[1]{\renewcommand{\baselinestretch}{#1}\large\normalsize}

\newcommand {\figureStretch}{\Stretch{0.9}}

\newcommand {\unStretch}{\Stretch{2}}

\begin{document}

\pagestyle{empty}

\Stretch{3}
\begin{center}

{\large 3/10/95}
\vspace*{2em}

{\huge \bf The density matrix renormalisation group and critical phenomena}
\vspace*{5em}

\Stretch{1.5}
{\large
R.\ J.\ Bursill and F.\ Gode
}
\vspace*{2em}

Department of Physics,
The University of Sheffield,
S3 7RH, Sheffield, United Kingdom.\\
Email: r.bursill@sheffield.ac.uk

\vspace*{2em}
Short Title -- DMRG and critical phenomena\\
Classification Number -- 05.30 -- Statistical Mechanics---Quantum\\
%Keywords -- quantum spin chain, renormalisation group, critical
%phenomena
\end{center}

\vfill
\eject

\Stretch{2}

\begin{abstract}

We adapt White's density matrix renormalisation group (DMRG) to the direct
study of critical phenomena. We use the DMRG to generate transformations in the
space of coupling constants. We postulate that a study of density matrix
eigenvalues
leads to a natural identification of renormalised blocks, operators and
Hamiltonians. We apply the scheme to the phase transition in the anisotropic
spin-1/2 Heisenberg chain. In the simplest case where the two most probable
states in odd sized blocks are used to construct approximate renormalisation
group transformations, we find qualitative improvement upon the standard real
space renormalisation group method for the thermal exponent $\nu$.

\end{abstract}

\setcounter{page}{0}
\vfill
\eject

\pagestyle{plain}

\section{Introduction}
\label{Introduction}
\setcounter{equation}{0}

The advent of White's density matrix renormalisation group (DMRG) method
\cite{development} has lead to some very successful studies of
low-lying
excitations and static correlation functions in a number of one dimensional
quantum lattice systems \cite{applications}. The method represents a major
improvement upon its precursor the (conventional, Wilson) real space
renormalisation group (RSRG) method \cite{Wilson} which generally gives poor
or slowly converging results for these quantities.

Both methods are truncated basis expansions in that a target state (such as the
ground state) of a large lattice is built up from blocks of sites from which
only a few {\it important} states are retained. The key difference between the
two methods is the way in which the important states are determined. For the
RSRG states are retained on the basis of energy whereas in the DMRG states are
retained on the basis of how likely they are to be part of the target state
being investigated.

It is well documented that the DMRG method works best when the system being
studied is away from criticality---that is, when there is a substantial energy
gap, or when the generic correlation functions decay exponentially, with a
correlation length of only a few lattice spacings
\cite{development,criticality1,criticality2}. In fact, efforts to study zero
temperature phase transitions by using obvious approaches such as investigating
the divergence of the correlation length \cite{criticality1} or the vanishing
of
an order parameter \cite{criticality2} have failed because it is very difficult
to determine these quantities accurately near the critical point.

It is well known, however, that the RSRG can be used to generate
renormalisation
group transformations in the space of coupling constants \cite{Pfeuty}. Useful
qualitative and even accurate quantitative results can be obtained for the
positions of phase transitions and critical exponents \cite{Pfeuty,Sarker}.

A next step is to investigate as to whether the DMRG can be used to generate
such transformations and yield accurate phase diagrams and associated critical
exponents, especially in cases where the RSRG fails. That is, it would be
useful
to develop a DMRG scheme for coupling constant transformations which has the
same accuracy, systematic improvability and portability as the DMRG scheme
currently being used for the calculation of energies and correlation functions.
Such a scheme might be very useful if applied to the investigation of critical
phenomena in complex models such as coupled fermion chains.

Such an investigation has been carried out in a recent series of papers by
Drzewi\'{n}ski and coworkers \cite{Drzewinski}. Results from the RSRG were
compared directly with those obtained from its DMRG analogue. In studies
of anisotropic $XY$ models with transverse fields, it was generally concluded
that the DMRG has no special advantage over the RSRG in calculating critical
points and exponents. These studies cannot however be considered exhaustive.
Firstly, the blocks and superblocks used were small---only a few sites.
Secondly, the models did not afford the total $z$-spin as a good quantum
number.
This sometimes complicated the process of identifying important block states
with renormalised block spin variables.

In this paper we consider a DMRG approach to coupling constant transformations
which makes use of the fact that within DMRG algorithms we accurately calculate
important states for very large blocks by retaining large numbers of states at
each iteration. We apply this approach to a simple model, the anisotropic
Heisenberg model, where critical properties are very well known \cite{ahm}. The
model has been studied using the RSRG \cite{Spronken} with less than
encouraging
results. We find that the DMRG approach yields qualitative improvement in the
nature of the convergence of the thermal critical exponent $\nu$ as the
blocksize is increased.

In the following sections we briefly outline the DMRG algorithm and describe
how
it can be used to generate coupling constant transformations. We then present
our results for the anisotropic Heisenberg model and compare them with the RSRG
results of reference \cite{Spronken}. We then conclude with some remarks on
future directions.

\section{The DMRG and coupling constant transformations}

The DMRG algorithm has been described in great detail \cite{development} so we
will be brief in our description, concentrating on those points which are
relevant to generating renormalisation group transformations. We restrict
ourselves to the infinite lattice algorithm which we use in our calculations
and
we describe the algorithm in the context of the Heisenberg spin chain
\begin{equation}
{\cal H}
=
\sum_{i}
\left[
S_{i}^{z}S_{i+1}^{z} +
\gamma \left( S_{i}^{x}S_{i+1}^{x} + S_{i}^{y}S_{i+1}^{y} \right)
\right]
\label{H}
\end{equation}
where $S_{i}$ is the spin operator of spin $S$ for site $i$ on the chain and
$\gamma$ is the anisotropy.

The DMRG is an iterative, truncated basis procedure whereby a large chain
(or superblock) is built up from a small number of sites by adding a small
number of sites at a time. At each stage the superblock consists of system and
environment blocks (determined from previous iterations) in addition to a
small number of extra sites. Also determined from previous iterations are
the matrix elements of the block Hamiltonians and the {\em active} spin
operators (those on the sites at the end(s) of the blocks) with respect to a
truncated basis. Tensor products of the states of the system block, the
environment block and the extra sites are then formed to provide a truncated
basis for the superblock. The ground state $\left| \psi \right\rangle$ of the
superblock is determined by a sparse matrix diagonalisation algorithm.

Next, a basis for an augmented block, consisting of the system block and a
specified choice of the extra sites, is formed from tensor products of system
block and site states. The augmented block becomes the system block in the next
iteration. However, in order to keep the size of the superblock basis from
growing, the basis for the augmented block is truncated. We form a density
matrix by projecting $\left|\psi\right>\left<\psi\right|$ onto the augmented
block which we diagonalise with a dense matrix routine. We retain the {\em most
probable} eigenstates (those with the largest eigenvalues) of the density
matrix in order to form a truncated basis for the augmented block that is
around the same size as the system block basis. Matrix elements for the
Hamiltonian and active site operators are then updated.

The environment block used for the next iteration is usually chosen to be a
reflected version of the system block. The initial system and environment
blocks are chosen to be a small number of sites, usually one or two. The
accuracy and computer requirements of the scheme is fixed by $n\tsub{s}$, the
number of states retained per block (of good quantum numbers) at each
iteration.
$n\tsub{s}$ determines the truncation error, which is the sum of the
eigenvalues
of the density matrix corresponding to states which are shed in the truncation
process. The error in quantities such as the ground state energy scale linearly
with the truncation error \cite{development}.

Now, at each iteration we generate coupling constant transformations $T$ as
follows. After forming the system block of size $L$ we construct a lattice
${\cal L}$ of size $2L$ consisting of two system blocks. A (small) subset $B$
of
the system block truncated basis states is identified with a complete basis for
a renormalised block of spins of size $L'$ where $L'=L/b$ and $b$ is the
renormalisation factor. $B\otimes B$ is then clearly identifiable with a
complete basis for a lattice ${\cal L}'$ of size $2L'$. Next, the matrix
elements of ${\cal H}$, the Hamiltonian for ${\cal L}$, with respect to
$B\otimes B$ (which are readily formed from the matrix elements of the
block Hamiltonian and active site operator(s)) are identified with those of a
renormalised Hamiltonian ${\cal H}'$ defined on ${\cal L}'$. $T$ is then
defined
by
\begin{equation}
\gamma' = T(b|\gamma)
\label{T}
\end{equation}
where $\gamma$ and $\gamma'$ are set(s) of coupling constant(s) which define
the
Hamiltonians ${\cal H}$ and ${\cal H}'$ respectively.

To complete the prescription we must have a suitable method for choosing the
set
$B$---it's size and makeup---and hence the renormalisation factor $b$, together
with the type of renormalised block it is identified as a basis for. We must
then check that our identification is valid in that the Hamiltonian matrix
element identification can in fact be made such that ${\cal H}'$ and ${\cal
H}$,
in a loose sense, have the same symmetries. Ideally, we would like ${\cal H}'$
and ${\cal H}$ to be of the same form. The resulting system of linear equations
for the renormalised couplings $\gamma'$ is typically overdetermined. This
implies a consistency check of the identification.

As to the choice of $B$, just as the density matrix eigenvalues determine which
states are to be retained in forming a new system block, so too can we use the
spectrum of the density matrix in order to guide our choice as to the makeup of
$B$. That is, we choose $B$ from states that make up the bulk of the weight of
the density matrix whose eigenvalues sum to unity.

\subsection*{Choice of the basis $B$}

We commence with the simple example of an $S=1/2$ model with $\gamma=1$
(isotropic case). We use a periodic superblock of the form
\ldots-site-system-site-environment-\ldots and augment the system block with
both
sites. The strongest density matrix eigenvalues are listed in
table~\ref{eigenvalues1} for the case of $n\tsub{s}=20$. The initial system
block is a single site so the blocks always have an odd number of sites. The
strongest density matrix eigenvalues lie in the sector of the Hilbert space
where the good quantum number $S\tsub{T}^{z}\equiv\sum_{i}S_{i}^{z}$ has small
magnitude. We see that there is a pair of strong eigenvalues in the
$S\tsub{T}^{z}=\pm\frac{1}{2}$ sectors which make up the bulk of the weight of
the density matrix.

An obvious choice for $B$ then is the states corresponding to these
eigenvalues.
We identify them with the up and down $z$-spin states of a renormalised
spin-1/2
operator. That is, the system block is renormalised to a single site---$b=L$.
As
we shall see, the advantage of such a choice of $B$ is that ${\cal H}'$ is of
the same form as ${\cal H}$.

Next, in table~\ref{eigenvalues2} we list the strongest density matrix
eigenvalues for the case of an open ended superblock of the form
system-site-site-environment. We again augment the system with both sites. The
initial system block is a pair of sites so the blocks are always even. We see
that the simplest possible identification involves the four strongest
eigenvalues---a pair in the $S\tsub{T}^{z}=\pm 1$
sectors and a pair in the $S\tsub{T}^{z}=0$ sector. It is natural to form $B$
from the four states corresponding to these eigenvalues and to identify $B$ as
a
basis for a block of two spin-1/2 sites ie $b=L/2$. Following \cite{Jullien1}
we
identify the elements of $B$ with singlet and triplet states. That is, we
identify the $S\tsub{T}^{z}=\pm 1$ states with $|\uparrow\uparrow\rangle$ and
$|\downarrow\downarrow\rangle$ and the strong and weak eigenvalue (low and high
energy) $S\tsub{T}^{z}=0$ states with
$\frac{1}{\sqrt{2}}(|\uparrow\downarrow\rangle-|\downarrow\uparrow\rangle)$ and
$\frac{1}{\sqrt{2}}(|\uparrow\downarrow\rangle+|\downarrow\uparrow\rangle)$
respectively. It is found that with this identification ${\cal H}'$ has the
same
form as ${\cal H}$.

Finally, in table~\ref{eigenvalues3} we list the strongest density matrix
eigenvalues for a spin-1 model
using odd sized blocks, periodic boundary conditions, augmenting two sites at a
time. We see that, for small $L$, the bulk of the density matrix made up from
the three strongest eigenvalues which lie in the $S\tsub{T}^{z}=0$ and
$S\tsub{T}^{z}=\pm 1$ sectors. When the corresponding states are (naturally)
identified with the $|0\rangle$ and $|\pm 1\rangle$ states of a spin-1
operator,
the resulting ${\cal H}'$ has the same form as ${\cal H}$. However, as $L$ is
increased, the second largest eigenvalue in the $S\tsub{T}^{z}=0$ sector
rapidly becomes comparable to the other retained states' and the identification
of the renormalised block with a spin-1 site becomes inconsistent. It
becomes necessary to incorporate the corresponding state into $B$ which, as in
the previous example, is identified as a basis for a block of two spin-1/2
sites.

The renormalised $S=1/2$ Hamiltonian ${\cal H}'$ is not however of the form
\Ref{H}
in that there are second and third neighbour interactions present with
dimerisation.
This is consistent with White's
analytic mappings between the spin-1 chain and coupled spin-1/2 chains
\cite{s=1tocoupledchains}. That is, it becomes natural to identify the
renormalised block with a single segment of couple chains.

\subsection{Discussion---systematic improvability of transformations}

Now that we have described the procedure for generating coupling constant
transformations, we discuss the question of its accuracy and systematic
improvability. These issues are central to the success of the DMRG in
calculating energies and correlation functions. We consider how the algorithm
should be scaled in order to obtain the most accurate results.

Now, suppose we choose $B$ in the same way at each iteration eg for a spin-1/2
system with odd blocks we always form $B$ from two states---the most probable
states in the $S\tsub{T}^{z}=\pm\frac{1}{2}$ sectors. We have the guiding
principle that, as long as
we respect the basic symmetries of the Hamiltonian, then for a fixed basis size
$|B|$, the accuracy of the transformations should increase with each iteration,
with calculated quantities converging to their exact values as the blocksize is
increased.

This principle stems from the fact that the ratio of intrablock to interblock
components of the Hamiltonian decreases as the blocksize is increased and the
exact ground state can asymptotically be written as a suitably symmetrised
product of the $B$ states. The principle in fact appears to be borne out in the
case of the spin-1/2 transverse Ising (ITF) model where slow but systematic
convergence of critical exponents with blocksize is achieved within a RSRG
scheme whereby blocks of sizes 3, 5, 7, 9 and 11 are diagonalised and the basis
$B$ is formed from the states of lowest energy \cite{Jullien2}.

This principle as applied to estimates of critical exponents is however
non-rigorous in general and within the DMRG scheme there is likely to be a
limit to accuracy of infinite lattice results imposed by the finiteness of
$n\tsub{s}$. That is, the finiteness of $n\tsub{s}$ imposes restrictions on how
accurately we can determine important states of large blocks. In practice it
may
therefore be profitable to enlarge $B$ and hence $L'$. This may have the
undesirable side effect of ${\cal H}'$ lying in an enlarged space with more
interactions.

Finally, we describe how the approach considered here differs from that used in
\cite{Drzewinski}. In \cite{Drzewinski}, relatively small superblocks (up to 12
sites) are diagonalised exactly, reduced density matrices are then constructed
for small system blocks (either 3 or 4 sites) and the basis $B$ is formed from
retaining the 2(4) most probable states for 3(4) site system blocks. This is
the
direct DMRG analogue of the RSRG calculations \cite{Spronken}. In the approach
considered here we again use only a handful of states to generate the RG
transformation but many states are retained at each iteration for the purpose
of
constructing successive system blocks. That way we attempt to work with
accurately determined {\it important} states of large blocks. Also, the density
matrix eigenvalues are used in order to obtain a natural selection of the basis
$B$.

\section{Application to the spin-1/2 chain with anisotropy}

We now apply our method to the $S=1/2$ case of the Hamiltonian given
in equation \Ref{H}. This model is integrable
and many properties have been calculated exactly \cite{ahm}. We only consider
$\gamma\geq 0$ here. In this regime the model has two trivial, stable fixed
points. At $\gamma=0$ (Ising fixed point) the ground states are the classical
N\'{e}el states and there is a finite energy gap and long range
antiferromagnetic order. At $\gamma=\infty$ ($XY$ fixed point) the model is
equivalent to a spinless fermion gas \cite{LSM}, the spectrum is a gapless
continuum and there is no long range order. There is a phase transition at
$\gamma=\gamma\tsub{c}=1$ (isotropic, Heisenberg point). This critical point
separates the gapless phase from the gapped, ordered doublet phase.

The phase transition is pathological in that the critical exponents are either
zero or infinite. For example, the thermal exponent $\nu$, describing the
divergence of the correlation length $\xi$ is usually defined by
\begin{equation}
\xi
\sim
\left|\gamma-\gamma\tsub{c}\right|^{-\nu}
\mbox{ as }\gamma\rightarrow\gamma\tsub{c}
\end{equation}
However, for the $S=1/2$ model we have the exact result \cite{Spronken}
\begin{equation}
\xi
\sim
e^{ -A|\gamma-1|^{-1/2} }
\end{equation}
whence $\nu=\infty$.

As mentioned, the RSRG method has been applied to the $S=1/2$ model
\cite{Spronken}. Blocks of size $L=3$, 5, and 7 were used and the basis $B$ was
comprised of the two low energy states (in the $S\tsub{T}^{z}=\pm 1/2$ sectors)
and was identified with the up and down states of a spin-1/2 site. The
estimates
of $\nu$ so obtained were spurious in that $\nu\downarrow 2$ as
$L\rightarrow\infty$.

Here we calculate $\nu$ from RG transformations derived within the DMRG scheme
described above. We use the periodic superblock, augmenting two sites per
iteration with a single site as an initial system block (hence $L$ is always
odd). We choose $|B|=2$ ($L'=1$, $b=L$), identifying the most probable states
in
the $S\tsub{T}^{z}=\pm 1/2$ sectors, which we denote by
$|\uparrow\rangle\rangle$ and $|\downarrow\rangle\rangle$, with the up and down
states of a single spin-1/2 site. The renormalised Hamiltonian then has matrix
elements
\begin{equation}
\langle\sigma'\rho'|{\cal H}'|\sigma\rho\rangle
\equiv
\langle\langle\rho'|\otimes\langle\langle\sigma'|
{\cal H}
|\sigma\rangle\rangle\otimes|\rho\rangle\rangle
\end{equation}

${\cal H}'$ has sixteen matrix elements ie $B\otimes B$ has four
elements. However, the DMRG algorithm preserves certain symmetries of the
Hamiltonian at every iteration. From conservation of total $z$-spin we have
\begin{equation}
\langle\sigma'\rho'|{\cal H}'|\sigma\rho\rangle = 0
\mbox{ if }\sigma'+\rho'\neq\sigma+\rho
\end{equation}
{}From spin-flip symmetries we also have the following obvious relations
\begin{eqnarray}
\langle\uparrow\uparrow|{\cal H}'|\uparrow\uparrow\rangle
& = &
\langle\downarrow\downarrow|{\cal H}'|\downarrow\downarrow\rangle
\\
\langle\uparrow\downarrow|{\cal H}'|\uparrow\downarrow\rangle
& = &
\langle\downarrow\uparrow|{\cal H}'|\downarrow\uparrow\rangle
\\
\langle\uparrow\downarrow|{\cal H}'|\downarrow\uparrow\rangle
& = &
\langle\downarrow\uparrow|{\cal H}'|\uparrow\downarrow\rangle
\end{eqnarray}
These symmetries reduce the number of independent matrix elements to 3 ie
$\langle\uparrow\uparrow|{\cal H}'|\uparrow\uparrow\rangle$,
$\langle\downarrow\uparrow|{\cal H}'|\downarrow\uparrow\rangle$ and
$\langle\uparrow\downarrow|{\cal H}'|\downarrow\uparrow\rangle$.

The identification of ${\cal H}'$ is simple. We can write ${\cal H}'$ in the
same form as ${\cal H}$ (up to a scale factor $J'$ an additive constant $C'$)
viz
\begin{equation}
{\cal H}'
=
J'\left[
S_{1}^{z}S_{2}^{z} +
\gamma' \left( S_{1}^{x}S_{2}^{x} + S_{1}^{y}S_{2}^{y} \right)
\right]
+C'I
\label{H'}
\end{equation}
where $I$ denotes the identity operator and $J'$, $C'$ and $\gamma'$ are
uniquely determined from the three independent matrix elements. As mentioned,
in
general the identification is overdetermined.

It is found that the renormalisation group transformation $\gamma'=T(b|\gamma)$
converges quite rapidly and uniformly with $n\tsub{s}$. We plot the
transformation for various block sizes $b$ (or $L$) in Fig.~\ref{gamma'} in the
case of $n\tsub{s}=50$. We also include a plot of the RSRG result
\cite{Spronken} for $b=3$ viz
\begin{equation}
T(3|\gamma) = \frac{
16\gamma^{3}
}{
1+\sqrt{1+8\gamma^{2}}
}
\end{equation}
We see that the RSRG and DMRG results are almost identical for $b=3$. We shall
see, however, that a marked difference that emerges for larger $b$.

We see that the transformation has trivial fixed points at $\gamma=0$ and
$\gamma=\infty$ and the critical point $\gamma=\gamma\tsub{c}=1$ is recovered
as
an unstable fixed point. This is to be expected, again from symmetry
considerations ie ${\cal H}$ only has full rotational symmetry at
$\gamma=\gamma\tsub{c}$.

Note that the same argument applies in the spin-1 case. As mentioned, however,
it was found that the self-mapping for the $S=1$ case rapidly became
inconsistent as $b$ was increased. This is consistent with exact
diagonalisation
results \cite{spin1} where it is found that the $S=1$ model is not critical at
$\gamma=1$ but rather at $\gamma=1.167(7)$.

\subsection*{Results for the thermal exponent $\nu$}

Following \cite{Spronken} we use the following standard relation in order to
obtain the thermal exponent
\begin{equation}
\nu = \frac{ \log b }{ \log T'\left(b|\gamma\tsub{c}\right) }
\end{equation}
We plot the estimates of the exponent as a function blocksize $b$ in
Fig.~\ref{criticalexponent} for various values of $n\tsub{s}$. We also include
the RSRG results from \cite{Spronken}. We see that the DMRG results are
qualitatively more consistent with the exact result $\nu=\infty$ in that DMRG
result increases with $b$ whereas the RSRG result monotonically decreases with
$b$. Further, the
DMRG result may increase without bound as $b\rightarrow\infty$. We plot
$e^{\nu}$ versus $b$ for the $n\tsub{s}=50$ case in Fig.~\ref{expnu}. This
function appears to be linear for large $b$, consistent with $\nu$ diverging
logarithmically with $b$.

\section{Discussion}

In this paper we have presented a scheme for using the DMRG algorithm to
generate RG transformations in the space of coupling constants. The scheme
allows us to study critical phenomena directly using the DMRG method. The
scheme makes use of the fact that the {\em important} states of large blocks
are
accurately calculated within the DMRG algorithm. Also, a natural solution to
the
perennial problem of the choice of renormalised block and the identification of
the renormalised Hamiltonian is proposed in terms of the density matrix
spectrum. This allows us to distinguish the different universality classes of
the isotropic point in the anisotropic spin-$S$ Heisenberg model. For $S=1/2$,
the RG transformation is a self map with the isotropic point as a critical
point. For $S=1$ the model maps onto an $S=1/2$ model with dimerisation and
longer ranged interactions that can be identified with a coupled $S=1/2$ chain.

We have applied the method to the pathological phase transition in the $S=1/2$
model. Results for the thermal exponent $\nu$ are qualitatively better than the
comparable RSRG results. That is, the exact result is $\nu=\infty$. The RSRG
yields $\nu\downarrow 2$ as $b\rightarrow\infty$ (where $b$ is the blocksize)
whereas the DMRG has $\nu$ increasing with $b$, seemingly without bound.

We believe that the scheme described here may be useful in studying critical
phenomena in quasi-one dimensional systems with many states per unit cell such
as coupled spin or fermion chains, extended Hubbard-type models with a charge
transfer gap and spin or fermion models with dimerisation and frustration.

The authors are grateful for valuable discussions with Prof. G.\ Gehring of
Sheffield University and Dr T.\ Xiang from the IRC in Superconductivity at
Cambridge. The programmes used for the calculations were adapted from a
programme written by T.\ Xiang. R.\ J.\ B.\ gratefully acknowledges the support
of SERC grant no. GR/J26748. F.\ Gode gratefully acknowledges the support of
the a Postgraduate Scholarship from the Government of Turkey.

\vfill
\eject

\section*{Table Captions}

\noindent 1. Strongest density matrix eigenvalues for the isotropic $S=1/2$
model using $n\tsub{s}=20$ for various odd blocksizes $b$ using periodic
boundary conditions and augmenting by 2 sites per iteration.

\noindent 2. Strongest density matrix eigenvalues for the isotropic $S=1/2$
model using $n\tsub{s}=20$ for various even blocksizes $b$ using open boundary
conditions and augmenting by 2 sites per iteration.

\noindent 3. Strongest density matrix eigenvalues for the isotropic $S=1$ model
using $n\tsub{s}=20$ for various odd blocksizes $b$ using periodic boundary
conditions and augmenting by 2 sites per iteration.

\vfill
\eject

\setcounter{table}{0}

\begin{table}[htbp]
\centering

\begin{tabular}{||c|c|l||}
\hline
$b$ & $S\tsub{T}^{z}$ & Strongest Eigenvalues \\
\hline
\hline
3  & $\pm 3/2$ & $0.0000$, $\ldots$ \\
\hline
3  & $\pm 1/2$ & $0.5000$, $0.0000$, $0.0000$, $\ldots$ \\
\hline
\hline
5  & $\pm 3/2$ & $0.0055$, $\ldots$ \\
\hline
5  & $\pm 1/2$ & $0.4619$, $0.0271$, $0.0055$, $\ldots$ \\
\hline
\hline
9  & $\pm 3/2$ & $0.0120$, $\ldots$ \\
\hline
9  & $\pm 1/2$ & $0.4251$, $0.0484$, $0.0120$, $\ldots$ \\
\hline
\hline
21 & $\pm 3/2$ & $0.0237$, $\ldots$ \\
\hline
21 & $\pm 1/2$ & $0.3786$, $0.0682$, $0.0237$, $\ldots$ \\
\hline
\end{tabular}

\caption{}
\label{eigenvalues1}
\end{table}

\begin{table}[htbp]
\centering

\begin{tabular}{||c|c|l||}
\hline
$b$ & $S\tsub{T}^{z}$ & Strongest Eigenvalues \\
\hline
\hline
4  & $\pm 1$ & $0.02752$, $0.00000$, $\ldots$ \\
\hline
4  & $0$ & $0.91746$, $0.02752$, $0.00000$ $\ldots$ \\
\hline
\hline
6  & $\pm 1$ & $0.03917$, $0.00003$, $\ldots$ \\
\hline
6  & $0$ & $0.88237$, $0.03917$, $0.00003$, $\ldots$ \\
\hline
\hline
10 & $\pm 1$ & $0.05204$, $0.00022$, $\ldots$ \\
\hline
10 & $0$ & $0.84310$, $0.05204$, $0.00022$, $\ldots$ \\
\hline
\hline
22 & $\pm 1$ & $0.06938$, $0.00104$, $\ldots$ \\
\hline
22 & $0$ & $0.78808$, $0.06938$, $0.00104$, $\ldots$ \\
\hline
\end{tabular}

\caption{}
\label{eigenvalues2}
\end{table}

\begin{table}[htbp]
\centering

\begin{tabular}{||c|c|l||}
\hline
$b$ & $S\tsub{T}^{z}$ & Strongest Eigenvalues \\
\hline
\hline
3  & $\pm 2$ & $0.00000$, $\ldots$ \\
\hline
3  & $\pm 1$ & $0.33333$, $0.00000$, $\ldots$ \\
\hline
3  & $0$ & $0.33333$, $0.00000$, $0.00000$, $\ldots$ \\
\hline
\hline
5  & $\pm 2$ & $0.00520$, $\ldots$ \\
\hline
5  & $\pm 1$ & $0.29495$, $0.00520$, $\ldots$ \\
\hline
5  & $0$ & $0.29495$, $0.07181$, $0.00520$, $\ldots$ \\
\hline
\hline
9  & $\pm 2$ & $0.00579$, $\ldots$ \\
\hline
9  & $\pm 1$ & $0.26230$, $0.00579$, $\ldots$ \\
\hline
9  & $0$ & $0.26230$, $0.15410$, $0.00579$, $\ldots$ \\
\hline
\hline
21 & $\pm 2$ & $0.00388$, $\ldots$ \\
\hline
21 & $\pm 1$ & $0.23786$, $0.00389$, $\ldots$ \\
\hline
21 & $0$ & $0.23786$, $0.22696$, $0.00389$, $\ldots$ \\
\hline
\hline
\end{tabular}

\caption{}
\label{eigenvalues3}
\end{table}

\vfill
\eject

\section*{Figure Captions}

\noindent 1. The renormalisation group transformation $\gamma'=T(b|\gamma)$ for
the anisotropic spin-1/2 chain from the DMRG method with $n\tsub{s}=50$ for
various blocksizes $b$. The thick dashed curve is the RSRG result for $b=3$.

\noindent 2. DMRG estimates of the critical exponent $\nu$ as a function of
blocksize $b$ for $n\tsub{s}=20$ (dashed line), $n\tsub{s}=35$ (dot-dashed
line)
and $n\tsub{s}=50$ (solid line). The thick dashed curve is the RSRG result.

\noindent 3. $e^{\nu}$ versus $b$ for the $n\tsub{s}=50$ case.

\vfill
\eject

\setcounter{figure}{0}

\figureStretch
\begin{figure}[htbp]
%\epsfboxtw{fig1.ill}
\caption{}
\label{gamma'}
\end{figure}
\unStretch

\figureStretch
\begin{figure}[hbtp]
%\epsfboxtw{fig2.ill}
\caption{}
\label{criticalexponent}
\end{figure}
\unStretch

\figureStretch
\begin{figure}[hbtp]
%\epsfboxtw{fig3.ill}
\caption{}
\label{expnu}
\end{figure}
\unStretch

\end{document}